\documentclass[runningheads,a4paper]{llncs}

\usepackage[english]{babel}
\usepackage[latin1]{inputenc}
\usepackage{graphicx}
\usepackage{amsmath,amsfonts,amssymb}
\usepackage{latexsym}
\usepackage{mathrsfs}
\usepackage{stmaryrd}
\usepackage{url}
\usepackage{epsfig}

\usepackage{tikz}

\def\PER{{\rm PER}}

\def\NP{{\rm NP}}
\def\OR{{\rm OR}}
\def\F{{\rm F}}
\def\P{{\rm P}}
\def\FP{{\rm FP}}

\def\VP{{\rm VP}}
\def\VNP{{\rm VNP}}
\def\SAT{{\rm SAT}}
\def\VCP{{\rm VCP}}
\def\IP{{\rm IP}}
\def\IPP{{\rm IPP}}
\def\AP{{\rm AP}}
\def\deg{{\rm deg \kern+2pt}}
\def\sharpP{{\rm \# P}}

\newcommand\zz{\ensuremath{\mathbb{Z}}}

\usepackage{url}
\urldef{\mails}\path|{irenee.briquel, pascal.koiran}@ens-lyon.fr|
\newcommand{\keywords}[1]{\par\addvspace\baselineskip
\noindent\keywordname\enspace\ignorespaces#1}

\begin{document}

\mainmatter  

\title{A Dichotomy Theorem for Polynomial Evaluation}


\author{Ir\'en\'ee Briquel\and Pascal Koiran}
%

\institute{LIP\thanks{UMR 5668 ENS Lyon, CNRS, UCBL associ\'ee à l'INRIA.},
\'Ecole Normale Sup\'erieure de Lyon, Universit\'e de Lyon\\
\mails}

\maketitle

\begin{abstract}
A dichotomy theorem for counting problems due to Creignou and Hermann states
that or any finite set $S$ of logical relations,
the counting problem $\# \SAT(S)$ is either in $\FP$, or
 $\sharpP$-complete.
In the present paper we show a dichotomy theorem for polynomial evaluation.
That is, we show that for a given set $S$, either there exists a $\VNP$-complete family of polynomials associated to $S$,
or the associated families of polynomials are all in $\VP$. We give a concise characterization of the sets $S$ that give rise to ``easy'' and ``hard'' polynomials.
We also prove that several problems which were known
to be $\sharpP$-complete under Turing reductions only
are in fact $\sharpP$-complete under many-one reductions.
\end{abstract}

\section{Introduction}.

In a seminal paper, Schaefer~\cite{SCH78} proved a dichotomy theorem
for boolean constraint satisfaction problems: he showed that for any finite
set $S$ of logical relations the satisfiability problem $\SAT(S)$ for $S$-formulas is either in $\P$, or $\NP$-complete.
Here, an $S$-formula over a set of $n$ variables
is a conjunction of relations of $S$ where the arguments
of each relation are freely chosen among the $n$ variables.
Schaefer's result was subsequently extended in a number of directions.
In particular, dichotomy theorems were obtained for counting problems,
optimization problems and the decision problem of
quantified boolean formulas.
An account of this line of work can be found in the book by Creignou, Khanna and
Sudan~\cite{CKS01}. In a different direction, constraint satisfaction
problems were also studied over non-boolean domains.
This turned out to be a surprisingly difficult question, and
it took a long time before a dichotomy
theorem over domains of size 3 could be obtained~\cite{bulatov}.

In the present paper we study polynomial evaluation  
from this dichotomic
point of view. We work within Valiant's algebraic framework: the role of
the complexity class $\NP$ in Schaefer's dichotomy theorem will be played
 by the class $\VNP$ of ``easily definable'' polynomial families, and
the role of $\P$ will be played by the class $\VP$ of ``easily computable''
polynomial families~\cite{Val79alg,Bur00}.
There is a well-known connection between counting problems and polynomial
evaluation. For instance, as shown by Valiant the permanent is complete
in both settings~\cite{Val79,Val79alg}.
In the realm of counting problems,
a dichotomy theorem was obtained by Creignou and Hermann~\cite{CH96,CKS01}.

\begin{theorem}
\label{countingDichotomy}
For any finite set $S$ of logical relations,
the counting problem $\# \SAT(S)$ is either in $\FP$, or
 $\sharpP$-complete.
\end{theorem}

In fact, the sets $S$ such that $\# \SAT(S)$ is in $\FP$ are exactly the sets containing only affine constraints (a constraint is called affine if it expressible as a system of linear equations over $\zz/2\zz$).

\subsection*{Main Contributions}

To a family of boolean formulas $(\phi_n)$ 
we associate the multilinear polynomial family
\begin{equation} \label{Pphi}
P(\phi_n)(\overline{X}) = \sum_{\overline{\varepsilon}}
\phi_n(\overline{\varepsilon}) \overline{X}^{\overline{\varepsilon}},
\end{equation}
where $\overline{X}^{\overline{\varepsilon}}$ is the monomial
$X_1^{\varepsilon_1} \cdots X_{k(n)}^{\varepsilon_{k(n)}}$,
and $k(n)$ is the number of variables of $\phi_n$.
Imagine that the $\phi_n$ are chosen among the $S$-formulas of a fixed finite set $S$
of logical relations. One would like to understand how the complexity
of the polynomials $P(\phi_n)$ depends on $S$.

\begin{definition}
A family $(\phi_n)$ of $S$-formulas is called a $p$-family
if $\phi_n$ is a conjunction of at most $p(n)$ relations from $S$,
for some polynomial $p$
(in particular, $\phi_n$ depends on polynomially many variables
when $S$ is finite).
\end{definition}

\begin{theorem}[Main Theorem] \label{bigtheorem}
Let $S$ be a finite set of logical relations. If $S$ contains only affine relations of at most two variables, then the families $(P(\phi_n))$ of polynomials associated to $p$-families of $S$-formulas $(\phi_n)$ are in $\VP$. Otherwise, there exists a $p$-family $(\phi_n)$ of $S$-formulas such that the corresponding polynomial family $P(\phi_n)$ is $\VNP$-complete.
\end{theorem}

We can remark, that the hard cases for counting problems are strictly included in our hard evaluation problems, exactly as the hard decision problems in Schaefer's theorem were strictly included in the hard counting problems.

In our algebraic framework the evaluation of the polynomial associated to a given formula consists in solving a ``weighted counting'' problem:
each assignment $(\varepsilon_1, \ldots, \varepsilon_k)$ of the variables of $\phi$ comes with a weight
$X_1^{\varepsilon_1}\cdots X_k^{\varepsilon_k}$. In particular, when the variables $X_i$ are all set to 1, we obtain
the counting problem $\# \SAT(S)$. It is therefore natural that evaluation problems turn out to be harder than counting problems.

The remainder of this paper is mostly devoted to the proof of
Theorem~\ref{bigtheorem}.

Allong the way, we obtain several results of independent interest. First, we obtain several new $\VNP$-completeness results.
The main ones are about:
\begin{itemize}
\item[(i)] the vertex cover polynomial $\VCP(G)$ and the independent set
polynomial $\IP(G)$, associated to a vertex-weighted graph $G$.
Most $\VNP$-completeness results in the literature (and certainly all the results in Chapter~3 of~\cite{Bur00}) are about edge-weighted
graphs.
\item[(ii)] the antichain polynomial $\AP(X)$ and the ideal polynomial $\IPP(X)$, associated to a weighted poset $(X,\leq)$. 
\end{itemize}

Unlike in most $\VNP$-completeness results, 
we need more general reductions to establish $\VNP$-completeness results than Valiant's $p$-projection.
In Section~\ref{affine}, we use the ``$c$-reductions'', introduced by Bürgisser~\cite{Bur99,Bur00} in his work on $\VNP$ families
that are neither $p$-computable nor $\VNP$-complete.
They are akin to the oracle (or Turing)
reductions from discrete complexity theory.
The $c$-reduction has not been used widely in $\VNP$-completeness proofs.
The only examples that we are aware of are:
\begin{itemize}
\item[(i)] A remark in~\cite{Bur00} on probability generating functions.
\item[(ii)] The  $\VNP$-completeness
of the weighted Tutte polynomial in~\cite{LoMa04}.
Even there, the power of $c$-reductions is used in a very restricted
way since a single oracle call is performed in each reduction.
\end{itemize}
By contrast, the power of oracle reductions has been put to good use
in $\sharpP$-completeness theory (mostly as a tool for performing
interpolation). Indeed, as pointed out in~\cite{Jer03}, ``interpolation
features prominently in a majority of $\sharpP$-completeness proofs'', and
``it is not clear whether the phenomenon of $\sharpP$-completeness would be
as ubiquitous if many-one reducibility were to be used in place of Turing.''
We argue that the importance of Turing reductions in  $\sharpP$-completeness
should be revised downwards since, as a byproduct of our
$\VNP$-completeness results, we can replace Turing reductions
by many-one reductions in several
$\sharpP$-completeness results from the literature.
In particular, we obtain a many-one version
of Creignou and Hermann's dichotomy theorem\footnote{Many-one reductions (Definition~\ref{def:manyone}) are called \emph{many-one counting reductions} in~\cite{CH96,CKS01}. It was already claimed in~\cite{CH96,CKS01} that Theorem~\ref{countingDichotomy} holds true for
many-one reductions. This was not fully justified since
the proof of Theorem~\ref{countingDichotomy} is based on many-one reductions
from problems which were previously known to be $\sharpP$-complete
under oracle reductions only. The present paper shows that this claim was
indeed correct.}.
We leave it as an open problem whether the 0/1 partial permanent is
$\sharpP$-complete under many-one reductions (see Section~\ref{monclauses}
for a definition of the partial permanent, and~\cite{Jerrum87} for a
\sharpP-completeness proof under oracle reductions).

\subsection*{Organization of the Paper and Additional Results}

Earlier in this section we gave an informal introduction
to constraint satisfaction problems.
We give more precise definitions at the beginning of Section~\ref{prelim}.
The remainder of that section is devoted to Valiant's algebraic model of
computation. We also deal briefly with the easy cases of Theorem~\ref{bigtheorem} (Remark~\ref{easycase}).
We then establish the proof of the hard cases of Theorem~\ref{bigtheorem}, beginning with the case of non affine constraints.
For that case, the high-level structure of the proof is similar to
Creignou and Hermann's proof of $\sharpP$-completeness
of the corresponding counting problems in~\cite{CH96}.
The singletons $S=\{\OR_2\}$, $S=\{\OR_1\}$ and $S=\{\OR_0\}$
play a special role in the proof.
Here $\OR_2$ denotes the negative two-clause
$(x,y)\mapsto (\overline{x}\vee \overline{y})$;
$\OR_0$ denotes the positive two-clause $(x,y)\mapsto (x\vee y)$;
and $\OR_1$ denotes the implicative two-clause
$(x,y)\rightarrow (\overline{x} \vee y)$.
The corresponding $\VNP$-completeness results for $S=\{\OR_2\}$ and $S=\{\OR_0\}$ are established in
section~\ref{monclauses}; the case of $\{\OR_1\}$ is treated in Section~\ref{implclauses}.
These results are put together with Creignou and Hermann's results in Section~\ref{general} to establish the existence of a $\VNP$-complete family for any set $S$ containing non affine clauses (Theorem~\ref{notaffine}). Section~\ref{affine} deals with the affine clauses with at least three variables (Theorem~\ref{theoaffine}).
This completes the proof of Theorem~\ref{bigtheorem}.
In Section~\ref{manyonesection}, we build on our $\VNP$-completeness results to prove
$\sharpP$-completeness under many-one reductions
for several problems which were only known to be $\sharpP$-complete under
oracle reductions.



\section{Preliminaries} \label{prelim}

\subsection{Constraint satisfaction problems}

We define a  logical relation to be a function from $\{0,1\}^k$ to $\{0,1\}$,
for some integer $k$ called the rank of the relation.
Let us fix a finite set $S=\{\phi_1, \dots, \phi_n\}$ of logical relations.
An $S$-formula over $n$ variables $(x_1, \ldots, x_n)$
is a conjunction of boolean formulas,
each of the form $g_i (x_{j_i(1)}, \dots, x_{j_i(k_i)})$
where each $g_i$ belongs to $S$ and $k_i$ is the rank of $g_i$.
In words, each element in the conjunction is obtained by applying
a function from $S$ to some variables chosen among the $n$ variables.

An instance of the problem $\SAT(S)$ studied by Schaefer~\cite{SCH78}
is an $S$-formula $\phi$, and one must decide whether $\phi$ is satisfiable.
For instance, consider the 3 boolean relations $\OR_0(x,y)=x\vee y$,
$\OR_1(x,y)= \overline{x} \vee y$ and
$\OR_2(x,y) = \overline{x}\vee \overline{y}$.
The classical problem 2-SAT is $\SAT(S)$ where $S=\{\OR_0,\OR_1,\OR_2\}$.
The counting problem $\# \SAT(S)$
was studied by Creignou and Hermann~\cite{CH96}.
In this paper we study the complexity of evaluating the polynomials $P(\phi)$
in~(\ref{Pphi}). We establish which sets $S$ give rise
to $\VNP$-complete polynomial families, and which one give rise only
to easy  to compute families.
We next define these notions precisely.

\subsection{$\sharpP$-completeness and $\VNP$-completeness}\label{sectionDefinitions}

Let us introduce the notion of many-one reduction for counting problems~\cite{Z91}:

\begin{definition}[Many-one reduction]\label{def:manyone}
Let $f : \{0,1\}^* \rightarrow \mathbb{N}$ and $g : \{0,1\}^* \rightarrow \mathbb{N}$ be two counting problems. A \emph{many-one reduction} from $f$ to $g$ consists of a pair of polynomial-time computable functions $\sigma : \{0,1\}^* \rightarrow \{0,1\}^*$ and $\tau : \mathbb{N} \rightarrow \mathbb{N}$ such that for every $x \in \{0,1\}^*$, the equality $f(x) = \tau ( g( \sigma (x)))$ holds. When $\tau$ is the identity function, this reduction is called \emph{parsimonious}.
\end{definition}
A counting problem $f$ is $\sharpP$-hard under many-one reduction if every problem in $\sharpP$ admits a many-one reduction to $f$.

In Valiant's model one studies the computation of multivariate polynomials.
This can be done over any field. In the sequel we fix a field $K$
of characteristic~$\neq 2$. All considered polynomials are over $K$.

A $p$-family is a sequence $f = (f_n)$ of multivariate polynomials such that the number of variables and the degree are 
polynomially bounded functions of $n$.
A prominent example of a  $p$-family is the permanent family $\PER = (\PER_n)$, where $\PER_n$ is the permanent of an $n \times n$ matrix
with independent indeterminate entries.

We define the complexity of a polynomial $f$ to be the minimum number
$L(f)$ of nodes of an arithmetic circuit computing $f$.
We recall that the internal nodes of an arithmetic circuit perform
additions or multiplications, and each input  node is labeled by
a constant from $K$ or a variable $X_i$.

\begin{definition}[VP]
A $p$-family $(f_n)$ is $p$-computable if $L(f_n)$ is a polynomially bounded
function of $n$.
Those families constitute the complexity class $\VP$.
\end{definition}
In Valiant's model, $\VNP$ is the analogue of the class $\NP$ (or perhaps more accurately, of $\sharpP$).
\begin{definition}[VNP]
A $p$-family $(f_n)$ is called $p$-definable if there exists a $p$-computable family $g=(g_n)$ such that
$$
f_n(X_1, \dots, X_{p(n)})= \sum_{\varepsilon \in \{0,1\}^{q(n)}} g_n (X_1, \dots, X_{p(n)}, \varepsilon_1, \dots, \varepsilon_{q(n)})
$$
The set of $p$-definable families forms the class $\VNP$.
\end{definition}

Clearly, $\VP$ is included in $\VNP$.
To define $\VNP$-completeness we need a notion of reduction:
\begin{definition}[$p$-projection]
A polynomial $f$ with $v$ arguments is said to be a projection of a polynomial $g$ with $u$ arguments, and we denote it $f \leq g$,
if $f(X_1,\dots, X_{v}) = g(a_1, \dots , a_u)$ where each $a_i$
is a variable of $f$  or a constant from $K$.

A $p$-family $(f_n)$ is a $p$-projection of $(g_m)$ if there exists a
polynomially bounded function $t: \mathbb{N} \rightarrow \mathbb{N}$ such that: $
\exists n_0 \forall n \geq n_0,   f_n \leq g_{t(n)}
$.
\end{definition}

\begin{definition}[VNP-completeness]
A $p$-family $g \in \VNP$ is $\VNP$-complete if every
$p$-family $f \in \VNP$ is a $p$-projection of $g$.
\end{definition}

The $\VNP$-completeness of the permanent
under $p$-projections~\cite{Val79alg,Bur00}
is a central result in Valiant's theory.

It seems that $p$-projections are too weak for some
of our completeness results. Instead, we use the more general notion of
$c$-reduction~\cite{Bur99,Bur00}.
First we recall the notion of oracle computation :
\begin{definition}
The \emph{oracle complexity} $L^g(f)$ of a polynomial $f$ with respect to the oracle polynomial $g$ is the minimum number of arithmetic operations ($+,*$)
 and evaluations of $g$ over previously computed values that are sufficient to compute $f$ from the indeterminates $X_i$ and constants from $K$.
\end{definition}

\begin{definition}[$c$-reduction]
Let us consider two p-families $f=(f_n)$ and $g=(g_n)$. We have a polynomial oracle reduction, or $c$-reduction, from $f$ to $g$ (denoted $f \leq_c g$)
if there exists a polynomially bounded function
$t : \mathbb{N} \rightarrow \mathbb{N}$
such that the map $n \mapsto L^{g_{t(n)}}(f_n)$ is polynomially bounded.
\end{definition}

We can define a more general notion of $\VNP$-completeness
based on $c$-reductions:
A $p$-family $f$ is $\VNP$-hard if $g \leq_c f$ for every $p$-family
$g \in \VNP$.  It is $\VNP$-complete if in addition, $f \in \VNP$.
The new class of $\VNP$-complete families contains all the classical
$\VNP$-complete families since every $p$-reduction is a $c$-reduction.

In our completeness proofs we need $c$-reductions to compute the homogeneous
components of a polynomial. This can be achieved thanks to a well-known
lemma (see e.g.~\cite{Bur00}):
\begin{lemma} \label{homo}
Let $f$ be a polynomial in the variables $X_1, \dots , X_n$. For any $\delta$ such that $\delta \leq \deg f$, let denote $f^{(\delta)}$ the homogeneous component of degree $\delta$ of $f$. Then, $L^f(f^{(\delta)})$ is polynomially bounded in the degree of $f$.
\end{lemma}

By Valiant's criterion (Proposition~2.20 in~\cite{Bur00}),
for any finite set $S$ of logical relations and
any $p$-family $(\phi_n)$ of $S$-formulas the polynomials $(P(\phi_n))$
form a $\VNP$ family. Furthermore, the only four boolean affine relations with at most two variables are $(x=0)$, $(x=1)$, $(x = y)$ and $(x \neq y)$. Since for a conjunction of such relations, the variables are either independent or completely bounded, a polynomial associated to a $p$-family of such formulas is immediately factorizable. Thus :

\begin{remark}\label{easycase}
For a set $S$ of affine relations with at most two variables, every $p$-family of polynomials associated to $S$-formulas is in $\VP$.
\end{remark}

All the work in the proof of Theorem~\ref{bigtheorem} therefore goes into the hardness proof.

\section{Monotone 2-clauses}\label{monclauses}

In this section we consider the set $\{ \OR_2\} = \{(x,y)\mapsto (\overline{x}\vee \overline{y})\}$ and $\{ \OR_0\} = \{(x,y)\mapsto ({x}\vee {y})\}.$
For $S=\{ \OR_2\}$ and $S=\{ \OR_0\}$,
we show that there exists a $\VNP$-complete family of polynomials $(P(\phi_n))$
associated to a $p$-family of $S$-formulas $(\phi_n)$.

The partial permanent $\PER^*(A)$ of a matrix $A=(A_{i,j})$ is defined
by the formula:
$$
\PER^*(A) = \sum_{\pi} \prod_{i \in \text{def} \pi} A_{i \pi (i)}
$$
where the sum runs over all injective partial maps from $[1,n]$ to $[1,n]$. It is shown in~\cite{Bur00} that the partial permanent is
$\VNP$-complete (the proof is attributed to Jerrum).
The partial permanent may be written as in~(\ref{Pphi}),
where $\phi_n$ is the boolean formula that
recognizes the matrices of partial maps from $[1,n]$ to $[1,n]$.
But $\phi_n$ is a $p$-family of $\{\OR_2\}$-formulas since
$$
\phi_n (\varepsilon) = \bigwedge_{i,j, k: j \neq k} \overline{\varepsilon_{ij}} \vee \overline{\varepsilon_{ik}} \wedge \bigwedge_{i,j,k: i \neq k} \overline{\varepsilon_{ij}} \vee \overline{\varepsilon_{kj}}.$$
Here the first conjunction ensures that the matrix $\varepsilon$ has no more than one 1 on each row; the second one ensures that $\varepsilon$ has no more than one 1 on each column.
We have obtained the following result.
\begin{theorem}
The family $(\phi_n)$ is a $p$-family of $\{ \OR_2\}$-formulas,
and the polynomial family $(P(\phi_n))$ is $\VNP$-complete
under $p$-projections.
\end{theorem}


The remainder of this section is devoted to the set
$S = \{\OR_0\}=\{(x,y) \mapsto x \vee y\}$.
The role played by the partial permanent in the previous section will
be played by vertex cover polynomials.
There is more work to do because the corresponding $\VNP$-completeness result
is not available from the literature.

Consider a vertex-weighted  graph $G= (V,E)$:
to each   vertex $v_i \in V$ is associated a weight $X_i$.
The vertex cover polynomial of $G$ is
\begin{equation} \label{vcp_def}
\VCP(G) = \sum_S \prod_{v_i \in S} X_i
\end{equation}
where the sum runs over all vertex covers of $G$ (recall that
a vertex cover of $G$ is a set $S \subseteq V$ such that
for each edge $e \in E$, at least one of the two endpoints of $e$ belongs to $S$).
The univariate vertex cover polynomial defined in~\cite{DongHTL02}
is a specialization of ours; it is obtained from $\VCP(G)$
by applying the substitutions $X_i:=X$ (for $i=1,\ldots,n$),
 where $X$ is a new indeterminate.

Our main result regarding $\{\OR_0\}$-formulas is as follows.
\begin{theorem} \label{vcp}
There exists a family $G_n$ of polynomial size bipartite graphs such that:
\begin{enumerate}
 \item The family $(\VCP(G_n))$ is $\VNP$-complete.
 \item $\VCP(G_n)=P(\phi_n)$ where $\phi_n$
is a $p$-family of $\{\OR_0\}$-formulas.
\end{enumerate}
\end{theorem}
Given a vertex-weighted graph $G$, let us  associate to each $v_i \in V$ a boolean variable $\varepsilon_i$. The interpretation is that $v_i$ is chosen
in a vertex cover when $\varepsilon_i$ is set to 1.
We then have
$$
\VCP(G) = \sum_{\varepsilon \in \{0,1\}^{|V|}} \big[ \bigwedge_{(v_i,v_j) \in E} \varepsilon_i \vee \varepsilon_j  \big] \overline{X}^{\overline{\varepsilon}}
.$$
The second property in Theorem~\ref{vcp} will therefore hold true for
any family $(G_n)$ of polynomial size graphs.

To obtain the first property, we first establish a $\VNP$-completeness
result for the independent set polynomial $\IP(G)$.
This polynomial is defined like the vertex cover polynomial,
except that the sum in~(\ref{vcp_def}) now runs over all independent sets $S$
(recall that
an independent set is a set $S \subseteq V$ such that
there are no edges between any two elements of $S$).
\begin{theorem} \label{ip}
There exists a family $(G'_n)$ of polynomial size graphs
such that $\IP(G'_n)=\PER^*_n$
where $\PER^*_n$ is the $n \times n$ partial permanent.
The family $\IP(G'_n)$ is therefore $\VNP$-complete.
\end{theorem}
\begin{proof}
The vertices of $G'_n$ are the $n^2$ edges $ij$ of the complete bipartite
graph $K_{n,n}$, and the associated weight is the indeterminate $X_{ij}$.
 Two vertices of $G'_n$ are connected by an edge if they share an
endpoint in $K_{n,n}$.
An independent set in $G'_n$ is nothing but a partial matching in $K_{n,n}$,
and the corresponding  weights are the same.
\end{proof}
Next we obtain a reduction from the independent set polynomial to
the vertex cover polynomial. The connection between these two problems is
not astonishing since vertex covers are exactly the complements
of independent sets.
But we deal here with weighted counting problems, so that there is
a little more work to do.
The connection between independent sets and vertex covers
does imply a relation between the polynomials  $\IP(G)$  and $\VCP(G)$.
Namely,
\begin{equation} \label{complement}
\IP(G)(X_1,\dots,X_n) = X_1 \cdots  X_n \cdot
\VCP(G)(1/X_1,\dots,1/X_n).
\end{equation}
Indeed,
$$\begin{array}{rcccl}
\IP(G) &
= & \displaystyle \sum_{S\ {\rm independent}} {X_1 \cdots  X_n \over
\prod_{v_i {\not \in} S} X_i}
&  = & X_1 \cdots  X_n
\displaystyle  \sum_{S'\ {\rm vertex\ cover}} {1 \over
\prod_{v_i {\in} S'} X_i}.
\end{array}$$

Recall that the incidence graph of a graph $G'=(V',E')$ is a bipartite
graph $G=(V,E)$ where $V=V' \cup E'$. In the incidence graph
there is an edge between $e' \in E'$ and $u' \in V'$ if $u'$
is one of the two endpoints of $e'$ in $G$.
When $G'$ is vertex weighted, we assign to each $V'$-vertex of $G$ the same
weight as in $G$ and we assign to each $E'$-vertex of $G$ the constant weight
$-1$.
\begin{lemma} \label{ipvcp}
Let $G'$ be a vertex weighted graph and $G$ its vertex weighted incidence
graph as defined above.
We have the following equalities: \begin{equation} \label{iptovcp}
\VCP(G)=(-1)^{e(G')}\IP(G')
\end{equation}
\begin{equation} \label{vcptoip}
\IP(G)=(-1)^{e(G')}\VCP(G')
\end{equation}
 where $e(G')$ is the number of edges
of $G'$.
\end{lemma}
\begin{proof}
We begin with~(\ref{iptovcp}). To each independent set $I'$ of $G'$
we can injectively associate the vertex cover $C=I' \cup E'$. The weight of $C$
is equal to $(-1)^{e(G')}$ times the weight of $I'$.
Moreover, the weights of all other vertex covers of $G$
 add up to 0.
Indeed, any vertex cover $C$ which is not of this form must contain
two vertices $u',v' \in V'$ such that $u'v' \in E'$.
The symmetric difference $C \Delta \{u'v'\}$ remains a vertex cover of $G$,
and its weight is opposite to the weight of $C$ since it differs from $C$
only by a vertex $u'v'$ of weight $-1$.

The equality~(\ref{vcptoip}) follow from the combination of~(\ref{complement}) and~(\ref{iptovcp}).
\end{proof}
To complete the proof of Theorem~\ref{vcp} we apply Lemma~\ref{ipvcp} to
the graph $G'=G'_n$ of Theorem~\ref{ip}. The resulting graph $G=G_n$
satisfies $\VCP(G_n)=\IP(G'_n)=\PER^*_n$ since $G'_n$ has an even number of
edges: $e(G'_n)=n^2(n-1)$.

\section{Implicative 2-clauses}\label{implclauses}

Here we consider the set $S = \{\OR_1\} = \{(x,y)\rightarrow (x \vee \overline{y})\}$. Those logical relations are called implicative, because $x \vee \overline{y}$ is equivalent to $y \Rightarrow x$.
The $\sharpP$-completeness of $\# \SAT (S)$ was established by a chain
of reductions in~\cite{PB83} and~\cite{LI86}.
Here we will follow this chain of reductions to find
a $\VNP$-complete family associated to $S$-formulas.
These two articles show consecutively that the problems of counting the independent sets, the independent sets in a bipartite graph, the antichains in partial ordered sets (posets), the ideals in posets, and finally satisfaction of implicative 2-clauses are $\sharpP$-complete. We will start from the family $(G'_n)$ such that $\IP(G'_n) = \PER^*_n$, whose existence is stated in Theorem~\ref{ip}, and follow the reductions for the counting problems.

We first transform the family $(G'_n)$ into a family of bipartite graphs, without changing the independent set polynomials.
\begin{lemma}\label{ipbipartite}
There exists a family of bipartite graphs $(G''_n)$ such that $\IP(G''_n)= \PER^*_n$, the partial permanent of size $n \times n$.
\end{lemma}
\begin{proof}
By Lemma~\ref{ipvcp}, we know how to transform a graph $G$ into a bipartite graph $G'$ such that $\VCP(G')=(-1)^{e(G)}\IP(G)$ and $\IP(G')=(-1)^{e(G)}\VCP(G)$ where $e(G)$ is the number of edges of $G$. By applying this transformation one more time to $G'$, we obtain a graph $G''$ such that:

$$
\IP(G'')=(-1)^{e(G')}\VCP(G')= (-1)^{e(G)} \IP(G)
$$

Thus, the transformation of $G$ into $G''$ consists of the replacement of each edge $(u,v)$ in $G$ by the subgraph represented in Figure~\ref{bipartite}.

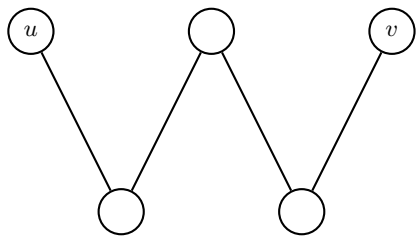
\begin{figure}
\begin{center}

      \begin{tikzpicture}[node distance=12mm]
        \tikzstyle{visible}=
        [%
          draw=black,%
          minimum size=6mm,%
          circle,%
          thick%
        ]

        \tikzstyle{invisible}=
        [%
        ]

        \node [visible] (A) {};
        \node [invisible] (B) [right of=A] {};
        \node [visible] (C) [right of=B] {$v$};

        \node [invisible] (D) [left of=A] {};
        \node [visible] (E) [left of=D] {$u$};
        \node [invisible] (F) [below of=A] {};
        \node [invisible] (G) [below of=F] {};
        \node [visible] (H) [left of=G] {};
        \node [visible] (I) [right of=G] {};

        \path [-,thick] (E) edge (H)
                         (H) edge (A)
                         (A) edge (I)
                         (I) edge (C);


      \end{tikzpicture}

\end{center}
\caption{The transformation of Lemma~\ref{ipbipartite}} \label{bipartite}

\end{figure}

In Theorem~\ref{ip} we introduced
a family $(G'_n)$ such that $G'_n$ has an even number of edges,
and $\IP(G'_n)=\PER^*_n$.
By applying the transformation above to  $(G'_n)$,
we obtain a bipartite
family $(G''_n)$ such that $\IP(G''_n)=\PER^*_n$.
\end{proof}
In the following we will not only use the statement of this lemma:
we will also use the structure of $G''_n$.
More precisely, let us denote by $V_1$ and $V_2$ the partite sets of $G''_n$.
We will use for instance the fact that in one of those two sets, say $V_1$, all vertices have weight $-1$.

It is pointed out in~\cite{PB83} that, given a bipartite graph,
one can construct naturally a partially ordered set. From the bipartite graph $G''_n=(V_1,V_2,E)$, we define the partially ordered set $(X_n, \leq)$ with $X_n = V_1 \cup V_2$, and given $x$ and $y$ in $X_n$, $x \leq y$ if and only if $x \in V_1$, $y \in V_2$ and $(x,y) \in E$. We see easily that $\leq$ is transitive and antisymmetric.

Next we recall the definition of an antichain.
\begin{definition}[Antichain]
An antichain $A$ in a poset $(X,\leq)$ is a subset of $X$ such that for all pair $(x,y)$ of distinct elements of $A$, $x$ and $y$ are incomparable.
\end{definition}
We define the antichain polynomial of a (weighted) poset $(X, \leq)$ as the polynomial:

$$
\AP(X) = \sum_{A} \prod_{x \in A} w(x)
$$
where the sum runs over all antichains $A$ of $(X, \leq)$.
Let us consider a bipartite graph $G$ and its corresponding poset $(X,\leq)$.
A set $S \subseteq X$ is an antichain in $(X, \leq)$ if and only if it is independent in $G$. We thus have: $\AP(X) = \IP(G)$.
Thus, we can identify the families $(\AP(X_n))$ and $(\IP(G''_n))$.
We then  define the notion of ideal in a poset.
\begin{definition}[Ideal]
An ideal $I$ in a poset $(X,\leq)$ is a subset of $X$ such that for all $x \in I$, all $y$ such that $y \leq x$ belong to $I$.
\end{definition}
We can also define the ideal polynomial $\IPP(X)$ in a poset $(X,\leq)$:
$$
\IPP(X) = \sum_{I} \prod_{x \in I} w(x)
$$
where the sum runs over all ideals $I$ of $(X,\leq)$.

Given an ideal $I$ in a poset $(X,\leq)$, the maximal elements of $I$ form an antichain $A$: since they are maximal in $I$, they cannot be compared. Conversely, given an antichain $A$, the set of elements $x$ that are not greater than an element of $A$ form an ideal.
One can verify easily that those transformations are bijective and inverse of each other. We thus have a bijection between the ideals and the antichains of a given poset. This fact suffices to the authors of~\cite{PB83}, since the bijection shows that a poset has the same number of antichains and ideals;
the counting problems are thus equivalent.

But for our weighted counting problems, since the ideals and the antichains have different weights, we cannot identify simply $\AP(X)$ and $\IPP(X)$ for any poset $X$. 

We do not know how to reduce a family of antichain polynomials into ideal polynomials in general, but in the case of the family $(\AP(X_n))$, since the structure of the family $(G''_n)$ is particular, the problem is easier. We claim the following.
\begin{theorem} \label{identity}
For all integers $n$, we have: $\AP(X_n) = \IPP(X_n)$
\end{theorem}
The proof will be given at the end of this section.
\begin{corollary}
There exists a $\VNP$-complete family of polynomials of the form $(\IPP(X_n))$.
\end{corollary}
To conclude, we note that the ideal polynomial in a poset $(X, \leq)$ may be expressed 
as a polynomial associated to a $S$-formula.
Namely, we associate to each $x_i \in X$ a boolean variable $\varepsilon_i$
with the intended meaning that $x_i$ belongs to an ideal when $\varepsilon_i$ is true. For every pair $(x_i,x_j) \in X$ such that $x_i \leq x_j$, the condition $ x_j \in I \Rightarrow x_i \in I$ may be expressed by $(\varepsilon_j \Rightarrow \varepsilon_i )$, or $(\varepsilon_i \vee \overline{\varepsilon_j})$. Thus, we have

$$
\IPP(X) = \sum_{\varepsilon \in \{0,1\}^{|X|}} \big[ \bigwedge_{(i,j): x_i \leq x_j} \varepsilon_i \vee \overline{\varepsilon_j} \big] \overline{X}^{\overline{\varepsilon}},$$
and as a result:
\begin{theorem}
There exists a $\VNP$-complete family of polynomials associated to a
$p$-family of $\{\OR_1\}$-formulas.
\end{theorem}
To complete this section, we now provide the proof of Theorem~\ref{identity}.
Let us fix an integer $n$. We recall that in the bipartite graph $G''_n=(V_1,V_2,E)$ constructed in Lemma~\ref{ipbipartite},
each vertex of $V_1$ has weight $-1$. We also know that $|V_1|$ is even, since the elements of $V_1$ are added two by two in the transformation from $G'_n$ to $G''_n$.

Fortunately, by modifying the correspondence between antichains and ideals, we can preserve the weights: we will construct in Lemma~\ref{lemmaBijection} a bijection from the antichains to the ideals of $X_n$ that preserves the weights, and thus we have:
$$\AP(X_n) = \IPP(X_n).$$
\begin{lemma}\label{lemmaBijection}
There exists a bijection (different from the natural one considered previously) from the antichains to the ideals of $X_n$, this one keeping the weights unchanged.
\end{lemma}
\begin{proof}
To an antichain $A$ of $X_n$, we associate the set $I$ such that:
\begin{itemize}
\item $A$ and $I$ coincide on $V_2$.

\item $I \cap V_1$ is the complement of $A \cap V_1$ in $V_1$.
\end{itemize}
The map $A \mapsto I$ is clearly injective,
and one can verify that the image $I$ is an ideal: given $x \in X_n$ and $y \in I$ such that $x \leq y$, we have that $x \in V_1$ and $y \in V_2$.
Therefore, $y \in A$, and $x$ cannot belong to $A$ as the elements of $A$ are incomparable. Thus, $x$ belong to $I$.
Our map is thus a bijection from the antichains to the ideals of $X_n$.

Since all the elements of $V_1$ have weight $-1$ and $|V_1|$ is even, the weights of $I$ and $A$ differ by a factor $(-1)^{|V_1|} = 1$.
\end{proof}

\section{Non affine constraints}
\label{general}

In this section we consider the general case of a set $S$ containing non affine constraints:

\begin{theorem}\label{notaffine}
For every set $S$ containing a non affine relation, there exists a $\VNP$-complete family (under $p$-projection) of polynomials associated to $S$-formulas.
\end{theorem}

The proof of this result is an analogue of the proof of the $\sharpP$-completeness of the corresponding counting problems given in~\cite{CKS01}. We will adapt this proof to our context.
The authors use the notion of perfect and faithful implementation (definition 5.1 in~\cite{CKS01}):

\begin{definition}
A conjunction of $\alpha$ boolean constraints $\{f_1, \dots, f_{\alpha}\}$ over a set of variables $\overline{x}=\{x_1, \dots, x_n\}$ and $\overline{y}=\{y_1, \dots, y_n\}$ is a \emph{perfect and faithful implementation} of a boolean formula $f(\overline{x})$, if and only if
\begin{enumerate}
 \item for any assignment of values to $\overline{x}$ such that $f(\overline{x})$ is true, there exists a unique assignment of values to $\overline{y}$ such that all the constraints $f_i(\overline{x},\overline{y})$ are satisfied.
 \item for any assignment of values to $\overline{x}$ such that $f(\overline{x})$ is false, no assignment of values to $\overline{y}$ can satisfy more than $(\alpha -1)$ constraints.
\end{enumerate}

We refer to the set $\overline{x}$ as the function variables
and to the set $\overline{y}$ as the auxiliary variables.
\end{definition}

We say, that a set $S$ of logical relations \emph{implements perfectly and faithfully} a boolean formula $f(\overline{x})$ if there is a $S$-formula that implements $f(\overline{x})$ perfectly and faithfully. We also extend the definition to logical relations: a set $S$ of logical relations implements perfectly and faithfully a logical relation $f$ if $S$ implements perfectly and faithfully every application of $f$ to a set of variables $\overline{x}$.

Let us denote $\F$ the unary relation $F(x)=\overline{x}$.
From~\cite{CKS01}, lemma 5.30, we have:

\begin{lemma}\label{lemmaImplementation}
If a logical relation $f$ is not affine, then $\{f,\F\}$ implements at least one of the three logical relations $\OR_0$, $\OR_1$ or $\OR_2$ perfectly and faithfully.
\end{lemma}

The following lemma, analogue to lemma 5.15 from~\cite{CKS01}, shows that perfect and faithful implementation provide a mechanism to do projections from the polynomials associated to sets of logical relations.

\begin{lemma}\label{lemmaProjection}
Let $S$ and $S'$ be two sets of logical relations such that every relation of $S$ can be perfectly and faithfully implemented by $S'$. Then every $p$-family of polynomials associated to a $p$-family of $S$-formulas is a projection of a $p$-family of polynomials associated to a $p$-family of $S'$-formulas.
\end{lemma}

\begin{proof}
Let $(\phi_n)$ be a $p$-family of $S$-formulas, and let us fix an integer $n$.

Let $\overline{x} = \{x_1, \dots , x_p\}$ be the set of variables of the formula $\phi_n$. This formula $\phi_n$ is a conjunction of logical relations $f_i \in S$ applied on variables from $\{x_1, \dots , x_p\}$.
If we replace each of those relations $f_i$ by a perfect and faithful implementation using constraints in $S'$, using for each $f_i$ a new set of auxiliary variables, we obtain a conjunction $\psi_n$ of logical relations from $S'$ applied on variable set $\overline{x} \cup \overline{y}$, where $\overline{y}= \{y_1, \dots, y_q\}$ is the union of the auxiliary variables sets added for each logical relation $f_i$.

Since all implementations are perfect and faithful, every assignment to $\overline{x}$ that satisfies all constraints of $\phi_n$ can be extended by a unique assignment to $\overline{x} \cup \overline {y}$ that satisfies all constraints of $\psi_n$. Conversely, for an assignment to $\overline{x}$ that does not satisfy all constraints of $\phi_n$, no assignment to $\overline{x} \cup \overline {y}$ can extend the previous one and satisfy every constraint of $\psi_n$.

Since $\psi_n$ is a conjunction of logical relations from $S'$ applied on a set of variables $\overline{x} \cup \overline {y}$, $\psi_n$ is a $S'$-formula. Furthermore, the number of constraints of $\psi_n$ is bounded by the product of the number of constraints of $\phi_n$ and the maximum number of logical relations from $S'$ needed to implement a logical relation from $S$ - which does not depend on $n$. The size of $\psi_n$ is therefore polynomially
linear in the size of $\phi_n$. We have:

\begin{eqnarray}
P(\phi_n) (X_1, \dots, X_p) & = & \sum_{\overline{\varepsilon} \in \{0,1\}^p} \phi_n(\overline{\varepsilon}) \overline{X}^{\overline{\varepsilon}} \nonumber\\
~ & = & \sum_{\overline{\varepsilon} \in \{0,1\}^p, \overline{y} \in \{0,1\}^q} \psi_n(\overline{\varepsilon}, \overline{y}) \overline{X}^{\overline{\varepsilon}} \nonumber\\
~ & = & \sum_{\overline{\varepsilon} \in \{0,1\}^p, \overline{y} \in \{0,1\}^q} \psi_n(\overline{\varepsilon}, \overline{y}) \overline{X}^{\overline{\varepsilon}} 1^{y_1} \dots 1^{y_q} \nonumber \\
~ & = & P(\psi_n) (X_1, \dots, X_p, 1, \dots, 1) \nonumber
\end{eqnarray}

Finally, the family $(P(\phi_n))$ is a projection of the family $(P(\psi_n))$, which is a $p$-family of polynomials associated to $S'$-formulas.
\end{proof}

From the two previous lemmas, and from the $\VNP$-completeness of families of polynomials associated to $\{\OR_0\}$- , $\{\OR_1\}$- and $\{\OR_2\}$-formulas, we conclude that for every set of logical relations $S$ such that $S$ contains non affine relations, there exists a $\VNP$-complete family of polynomials associated to $S \cup\{\F\}$-formulas.
To get rid of the logical relation $\{\F\}$, the authors of~\cite{CKS01} need to re-investigate the expressiveness of a non affine relation, and distinguish various cases. For our polynomial problems, we can easily force a boolean variable to be set to false by giving to the associated polynomial variable the value 0. We can now give the proof of Theorem~\ref{notaffine}:

\begin{proof}
Let $(\phi_n)$ be a $p$-family of $S \cup \{\F\}$-formulas such that $(P(\phi_n))$ is $\VNP$-complete. The existence of such a family is ensured by lemmas~\ref{lemmaImplementation} and~\ref{lemmaProjection}.

Let us consider an integer $n$. $\phi_n(x_1, \dots, x_n)$ is a conjunction of logical relations from $S$ applied to variables from $\overline{x}$ and and constraints of the form $(x_i = 0)$.
We remark, that if $\phi_n(\overline{x})$ contains the constraint $(x_i = 0)$, then the variable $X_i$ does not appear in the polynomial $P(\phi_n)(X_1, \dots, X_n)$: all the monomials containing the variable $X_i$ have null coefficients. If we suppress from the conjunction the constraint $(x_i = 0)$, and instead replace the corresponding variable $X_i$ by 0, we obtain exactly the same polynomial: the monomials such that $X_i$ appears in it have null coefficients; the others correspond to assignments such that $x_i=0$.

Let us denote $\psi_n$ the formula obtained by suppressing from $\phi_n$ all the constraints of the form $(x_i = 0)$. Since $P(\phi_n)(X_1, \dots, X_n) = P(\psi_n)(y_1, \dots, y_n)$, where $y_i$ is 0 if the constraint $(x_i = 0)$ was inserted in $\phi_n$, and $X_i$ otherwise, $(P(\phi_n))$ is a $p$-projection of $(P(\psi_n))$. Thus, the family $(P(\psi_n))$ is $\VNP$-complete.
\end{proof}

\section{Affine relations with at least three variables}\label{affine}

Here we consider the case of a set $S$ containing large affine constraints. We first establish the existence of a $\VNP$-complete family of polynomials associated to a $p$-family of affine formulas, and then show how to reduce this family to each affine constraint with at least three variables. In this section, our $\VNP$-completeness results are in the sense of $c$-reduction.

Let us consider the $n \times n$ permanent $\PER_n(M)$ of a matrix $M = (M_{i,j})$. It may be expressed as the polynomial associated to the formula accepting the $n \times n$ permutation matrices: $\PER_n (M) = \sum_{\varepsilon} \phi_n (\varepsilon) \overline{X}^{\overline{\varepsilon}}$

This formula $\phi_n$ expresses, that each row and each column of the matrix $\varepsilon$ contains exactly one 1. Let us consider the formula $\varphi_n$ defined by:

$$
\varphi_n(\varepsilon) = \bigwedge_{i=1}^{n} \varepsilon_{i1} \oplus \ldots \oplus \varepsilon_{in}=1 \wedge \bigwedge_{j=1}^{n} \varepsilon_{1j} \oplus \ldots \oplus \varepsilon_{nj} = 1
$$

The formula $\varphi_n$ expresses, that each row and each column of $\varepsilon$ contains an odd number of values 1. Thus, $\varphi_n$ accepts the permutation matrices, and other assignments that contain more values 1. We therefore remark, that the $n \times n$ permanent is exactly the homogeneous component of degree $n$ of $P(\varphi_n)$. But from Lemma~\ref{homo}, this implies a $c$-reduction from the permanent family to the $p$-family $(P(\varphi_n))$. Thus:

\begin{lemma}
The family $(P(\varphi_n))$ is $\VNP$-complete with respect to $c$-reductions.
\end{lemma}

Through $c$-reductions and $p$-projections, this suffices to establish the existence of $\VNP$-complete families for affine formulas of at least three variables:

\begin{theorem}\label{theoaffine}
\begin{enumerate}
\item There exists a $\VNP$-complete family of polynomials associated to $\{ x \oplus y \oplus z = 0 \}$-formulas.
\item There exists a $\VNP$-complete family of polynomials associated to $\{ x \oplus y \oplus z = 1 \}$-formulas.
\item For every set $S$ containing an affine formula with at least three variables, there exists a $\VNP$-complete family of polynomials associated to $S$-formulas.
\end{enumerate}
\end{theorem}

\begin{proof}

\begin{enumerate}

\item Let us consider the formula $\varphi_n$. This formula is a conjunction of affine relations with constant term 1: $x_1 + \ldots + x_k = 1$. Let $\varphi'_n$ be the formula obtained from $\varphi_n$ by adding a variable $a$ and replacing such clauses by $x_1 + \ldots + x_k + a = 0$. In the polynomial associated to $\varphi'_n$, the term of degree 1 in the variable associated to $a$ is exactly the polynomial $P(\varphi_n)$: when $a$ is assigned to 1, the satisfying assignments of $\varphi_n'$ are equal to the satisfying assignments of $\varphi_n$. Since this term of degree 1 can be recovered by polynomial interpolation of $P(\varphi'_n)$, the family $(P(\varphi_n))$ $c$-reduces to $(P(\varphi'_n))$.

 $\varphi'_n$ is a conjunction of affine relations with constant term 0. The polynomial $P(\varphi'_n)$ is the projection of the polynomial $P(\psi_n)$, where the formula $\psi_n$ is obtained from $\varphi'_n$ by replacing each affine relation of the type $x_{1} \oplus \ldots \oplus x_{k}=0$ by the conjunction of relations
$$
(x_1 \oplus x_2 \oplus a_1 = 0) \wedge (a_1 \oplus x_3 \oplus a_2 = 0) \wedge \ldots \wedge (a_{k-2} \oplus x_{k-1} \oplus x_{k} = 0)
$$
where the $a_i$ are new variables. In fact, on sees easily, that for a given assignment of the $x_i$ satisfying $\varphi'_n$, a single assignment of the $a_i$ gives a satisfying assignment of $\psi_n$; and that if the $x_i$ do not satisfy $\varphi'_n$, no assignment of the $a_i$ works on. The polynomial $P(\varphi'_n)$ is thus the polynomial obtained by replacing the variables associated to $a_i$ by the value 1 in $P(\psi_n)$; the family $(P(\varphi'_n))$ is a $p$-projection of $(P(\psi_n))$.

\item The formula $\psi_n$ constructed above is a conjunction of relations of the type $x \oplus y \oplus z = 0$. Let us construct a new formula $\psi'_n$ by introducing two new variables $a$ and $b$ and replacing each of such relations by the conjunction $(x \oplus y \oplus a = 1) \wedge (a \oplus z \oplus b= 1)$. One sees easily, that $P(\psi_n)$ is the projection of $P(\psi'_n)$ obtained by setting the variables associated to $a$ and $b$ to 1 and 0 respectively.

\item Let us suppose, that $S$ contains a relation of the type $x_1 \oplus \ldots \oplus x_k = 0$, with $k \geq 3$. The polynomial $P(\psi_n)$ is the projection of the polynomial associated to the $S$-formula obtained by replacing each relation $x \oplus y \oplus z = 0$ of $\psi_n$ by a relation $x \oplus y \oplus z \oplus a_1 \oplus \ldots \oplus a_{k-3} = 0$, and setting the variables associated to the $a_i$ to 0. Thus, the family $(P(\psi_n))$ projects on a family of polynomials associated to $S$-formulas, which is therefore $\VNP$-complete. When $S$ contains a relation with constant term 1, one projects the family $(P(\psi'_n))$ similarly.

\end{enumerate}

\end{proof}

\section{$\sharpP$-completeness proofs}
\label{manyonesection}

Up to now, we have studied vertex weighted graphs mostly from the point of
view of algebraic complexity theory. Putting weights on edges, or on vertices, can also be useful as an intermediate
step in $\sharpP$-completeness proofs~\cite{Val79,Jerrum87}.
 Here we follow this method to obtain new $\sharpP$-completeness
results. Namely, we prove $\sharpP$-completeness under many-one reductions
for several problems which were only known to be $\sharpP$-complete under
oracle reductions.

\begin{theorem} \label{manyone}
The following problems are $\sharpP$-complete for many-one reductions.
\begin{enumerate}
\item\label{itemVC} Vertex Cover:
counting the number of vertex covers of a given a graph.

\item\label{itemIS} Independent Set: counting the number of independent sets of a given
graph.

\item\label{itemBVC} Bipartite Vertex Cover:
 the restriction of vertex cover to bipartite graphs.

\item\label{itemBIS} Bipartite Independent Set:
 the restriction of independent set to bipartite graphs.

\item\label{itemA} Antichain: counting the number of antichains of a given poset.

\item\label{itemI} Ideal: counting the number of ideals of a given poset.

\item\label{itemI2sat} Implicative 2-$\SAT$: counting the number of satisfying assignments of a conjunction of implicative 2-clauses.

\item\label{itemP2sat} Positive 2-$\SAT$: counting the number of satisfying assignments of a conjunction of positive 2-clauses.

\item\label{itemN2sat} Negative 2-$\SAT$: counting the number of satisfying assignments of a conjunction of negative 2-clauses.
\end{enumerate}
\end{theorem}

\begin{remark}
$\sharpP$-completeness under oracle reductions is established
in~\cite{PB83} for the first six problems,
in~\cite{LI86} for the~\ref{itemI2sat}th problem and in~\cite{Val79c} for the last two.
In Section~\ref{prelim}, the last three problems are denoted $\# \SAT(S)$
where $S$ is respectively equal to $\{\OR_1\}$,  $\{\OR_0\}$ and $\{\OR_2\}$.
\end{remark}

\begin{proof}
Provan and Ball establish in~\cite{PB83} the equivalence of Problems~\ref{itemVC} and~\ref{itemIS}, \ref{itemBVC} and~\ref{itemBIS}, and~\ref{itemA} and~\ref{itemI}; they produce many-one reductions from~\ref{itemVC} to~\ref{itemP2sat} and from~\ref{itemBIS} to~\ref{itemA}, and Linial gives in~\cite{LI86} a many-one reduction from~\ref{itemI} to~\ref{itemI2sat}. Problems~\ref{itemP2sat} and~\ref{itemN2sat} are clearly equivalent. Therefore, to obtain $\sharpP$-completeness under many-one reductions for all those problems, we just need to show the $\sharpP$-completeness of Problem~\ref{itemVC} and to produce a many-one reduction from Problem~\ref{itemVC}
to Problem~\ref{itemBVC} (replacing the oracle reduction from~\cite{PB83}).

In order to prove the $\sharpP$-completeness of Problem~\ref{itemVC}, we first establish a many-one reduction from the $\sharpP$-complete problem of computing the permanent of $\{0,1\}$-matrices (which is known to be $\sharpP$-complete under many-one reductions~\cite{Z91}) to the problem of computing the vertex cover polynomial of a weighted graph with weights in $\{0,1,-1\}$.
    In~\cite{Bur00}, Bürgisser attributes to Jerrum a projection from the permanent to the partial permanent, with the use of the constant $-1$. Applied to a $\{0,1\}$-matrix, this gives a many-one reduction from the permanent on $\{0,1\}$-matrices to the partial permanent on $\{0,1,-1\}$-matrices.
    By  Theorem~\ref{ip}, the $n \times n$ partial permanent is equal to the independent set polynomial of the graph $G'_n$; the reduction is obviously polynomial. Moreover, by Lemma~\ref{ipvcp} this polynomial is the projection of the vertex cover polynomial of $G_n$, with the use of the constant $-1$. The partial permanent on entries in $\{0,1,-1\}$ therefore reduces
to the vertex cover polynomial on graphs with weights in $\{0,1,-1\}$.

    Let $G$ be such a vertex weighted graph, with weights in $\{0,1,-1\}$. A vertex cover of nonzero weight does not contain any vertex $v$ of weight $0$, and in order to cover the edges that are incident to $v$, it must contain all its neighbors. One can therefore remove $v$, and replace each edge from $v$ to another vertex $u$ by a self-loop (an edge from $u$ to~$u$). Thus, we obtain a graph $G'$ with weights in $\{1,-1\}$
such that $\VCP(G) = \VCP(G')$.

    To deal with the weights $-1$, we use a method similar to~\cite{Val79}. Since $\VCP(G')$ is the value of a partial permanent on a $\{0,1\}$-matrix, it is positive.
    We will construct an integer $N$  and a graph $H$
such that the number of vertex covers of $H$ modulo $N$ is equal to $\VCP(G')$.
This will establish a reduction from the boolean permanent to counting vertex covers.

We choose $N$ larger than the maximum value of the number of vertex covers
of $G'$: $N=2^{v(G')} +1$ will suit our purposes. Now that we compute the number of vertex covers modulo $N$, we can replace each $-1$ weight  in $G'$ by the weight $N-1=2^{v(G')}$. But one can simulate such a weight on a vertex by adding to it ${v(G')}$ leaves.

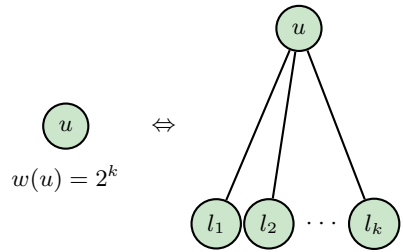
\begin{figure}
\begin{center}

      \begin{tikzpicture}[node distance=13mm]
        \tikzstyle{visible}=
        [%
          fill=green!50!black!20,%
          draw=black,%
          minimum size=6mm,%
          circle,%
          thick%
        ]

        \tikzstyle{invisible}=
        [%
          fill=white, 
          draw=white, 
          rectangle,%
          thick%
        ]

        \node [invisible] (A) {};
        \node [invisible] (F) [below of=A] {$\Leftrightarrow$};
        \node [invisible] (G) [below of=F] {};
        \node [visible] (H) [left of=F] {$u$};
        \node [invisible] (D) [below of=H, node distance = 7mm] {$w(u)=2^k$};

        \node [visible] (C) [right of=A, node distance = 18mm] {$u$};

        \node [visible] (I) [right of=G, node distance = 7mm] {$l_1$}
          edge[-, draw=black,thick]  (C);
        \node [visible] (J) [right of=I, node distance = 7mm] {$l_2$}
          edge[-, draw=black,thick]  (C);
        \node [invisible] (K) [right of=J, node distance = 7mm] {$\cdot \cdot \cdot$};

        \node [visible] (L) [right of=K, node distance = 7mm] {$l_k$}
          edge[-, draw=black,thick]  (C);

      \end{tikzpicture}

\end{center}

\caption{Simulation of a weight $2^k$} \label{weight}

\end{figure}

Finally, we construct a many-one reduction from vertex cover to bipartite vertex cover.
From Lemma~\ref{ipbipartite}, we have a projection from the
vertex cover polynomial of a graph to the vertex cover of a bipartite graph,
with the use of $-1$ weights.
To eliminate these weights, we can follow the method used in our above proof
of the $\sharpP$-completeness of Problem~\ref{itemVC}.
Indeed, since the leaves added to the graph preserve bipartiteness,
we obtain a reduction from counting vertex covers in a general graph
 to counting vertex covers in a bipartite graph.
\end{proof}
The proof of Creignou and Hermann's dichotomy theorem~\cite{CH96,CKS01}
is based on many-one reductions from the last 3 problems
of Theorem~\ref{manyone}.
We have just shown that these 3 problems are $\sharpP$-complete
under many-one reductions.
As a result, we have the following corollary to Theorem~\ref{manyone}.
\begin{corollary}
Theorem~\ref{countingDichotomy} still holds for $\sharpP$-completeness under many-one reduction.
\end{corollary}

\bibliography{briquel}

\end{document}